\documentclass[twocolumn,secnumarabic,amssymb, nobibnotes, aps, prd]{revtex4}

\usepackage{graphicx}

\begin{document}

\title{The origin of the Langevin equation and the calculation of the mean squared displacement: Let's set the record straight}%

\author{K. Razi Naqvi}%
\email[K. Razi Naqvi: ]{razi.naqvi@phys.ntnu.no}
\affiliation{Department of Physics, Norwegian University of Science and Technology (NTNU), N-7491 Trondheim, Norway}
\date{\today}%

\begin{abstract}
Ornstein and his coauthors, who constructed a dynamical theory of Brownian motion, taking the equation $mdv/dt =-\zeta v+X$ as their starting point, usually named the equation after Einstein alone or after both Einstein and Langevin; furthermore, Ornstein, who was the first to extract from this equation the correct expression for $\overline{\Delta^2}$, the mean-squared distance covered by a Brownian particle, credited de Haas-Lorentz, rather than Langevin, for finding the stationary limit of $\overline{\Delta^2}$. A glance at Einstein's 1907 paper, titled ``Theoretical remarks on Brownian motion'', should suffice to convince one that it is not unfair to attribute the {\it conception\/} of the above equation, now universally known as the Langevin equation, to Einstein. Langevin's avowed aim in his 1908 article was to recover, through a route that was `infinitely more simple', Einstein's 1905 expression for the diffusion coefficient, but a careful reading of Langevin's paper shows that---depending on how one interprets his description of the statistical behavior of the random force  $X$ appearing in the above equation---his analysis is at best incomplete, and at worst a mere tautology.  Since textbook accounts are based on the interpretation that renders the proof fallacious, alternative derivations, which are adaptations of those given by de Haas-Lorentz and Ornstein, are presented here. Some neglected aspects of the contents of Ornstein's early papers on Brownian motion are also brought to light. 

\end{abstract}

\maketitle

\section{Introduction}

It has become customary$^{1}$ to mark the birth of stochastic differential equations with the publication of the 1908 piece${}^{2,3}$ by Langevin---because he was the first to inscribe such an equation. It is equally common to regard Ornstein as the principal architect of the modern theory of Brownian motion---because he showed, in a series of seminal papers,${}^{4-9}$ how one may extract physically significant information from what is now called Langevin's equation. Yet it does not appear to be widely known that, in nearly all his papers, Ornstein named the underlying stochastic equation after Einstein alone or after both Einstein and Langevin. 

Though the question ``Did Ornstein have any grounds for associating the archetypal stochastic equation with Einstein?'' will be answered in this article, it will not occupy center-stage. My principal purpose is rather to draw attention to an unsatisfactory feature of Langevin's ostensibly simple derivation of an expression for the mean-squared displacement. One of his propositions (that relating to the vanishing of the average value of the product of the particle position $x$ and the random force $X$), on which the entire argument hinges, is offered without any proof; the singular ease with which the final result is arrived at provokes unease, rather than approval, when one notices that the average values of $vX$ and $xv$, two closely related products involving the velocity $v$,  cannot vanish, but they must do so according to the reasoning employed by Langevin.   

I argue here that Langevin's logic is seriously flawed, and its continued presentation to beginners no better than the dispensation of a nostrum to an unsuspecting patient. A demolition plan, such as that proposed here, stands a better chance of adoption if it is accompanied by a constructive suggestion; accordingly, I also present an alternative derivation, inspired by the edifice constructed by Ornstein and co-authors; though not as short as the original Langevin version, it is logically sound, and yields much greater insight. I also draw attention to an alternative treatment, due originally to de Haas-Lorentz,${}^{10}$ who may be called a forerunner of Ornstein, and to some unaccountably neglected aspects of Ornstein's work.

\section{Langevin, his forerunners, contemporaries and aftercomers}

Using different approaches, Einstein${}^{11}$ and Smoluchowski${}^{12}$ had arrived, shortly before Langevin turned his attention to the issue, at discordant expressions for the diffusion coefficient $D$ of a spherical particle. Langevin lodged two claims at the start of his analysis: ``I have been able to determine, first of all, that a correct application of the method of Mr. Smoluchowski leads one to recover the formula of Mr. Einstein {\it precisely\/}, and, furthermore, that it is easy to give a demonstration that is infinitely more simple by means of a method that is entirely different''. [Emphasis in the original]

It will be well to pause at this point and recall what Einstein wrote in his 1907 paper. I recapitulate, in the next two paragraphs, his statements after some condensation and a modicum of paraphrasing, but without changing his notation.

\subsection{Einstein's 1907 paper}

If we knew nothing of the molecular theory of heat, Einstein wrote,${}^{13}$ we should expect the following to happen. Suppose that we impart to a particle suspended in a liquid a certain velocity applied to it from without; then the particle will come to a halt as a result of the friction of the liquid. If we ignore the inertia of the latter and assume that the resistance experienced by a particle moving with velocity $v$ is $6\pi kPv$, where $k$ is the viscosity of the liquid and $P$ the radius of the particle, we obtain the equation
\begin{equation}
m{d v\over d t}= -6\pi kP v. 
\end{equation}
From this it follows that the velocity will fall to a tenth of its original value in a time $\theta =\ln 10\, m/(6\pi k P)$; for a spherical particle of radius 2.5 $\mu$m suspended in water, $\theta$ comes out to be 330 ns.

If we take into consideration the molecular theory of heat, Einstein reminded the reader, we have to modify this conception. We must continue to assume, as before, that the particle will nearly completely lose its initial velocity in the very short time $\theta$ through friction. But, at the same time, we must assume that the particle gets new impulses during this time by some process that is the inverse of viscosity, so that it retains a speed which on an average is equal to that implied by the Maxwell-Boltzmann distribution. But since we must imagine that direction and magnitude of these impulses are (approximately) independent of the initial velocity of the particle, we must conclude that the velocity of the particle will be very greatly altered in the extraordinary short time $\theta$, and, indeed, in a totally random manner. 

It is clear that Einstein conceived, already in 1907, a modified version of Eq.~(1), the modification entailing the  addition, on the right-hand side, of a random force. We shall see (in the next section) that his assumptions about the randomness of the additional force were echoed by Langevin when the latter put pen to paper. `Let us with Einstein consider a colloidal particle,' wrote Ornstein et al.$^9$ in 1927 (without citing a specific Einstein article); they presented in their own phrase the above argument from Einstein's 1907 note, and wrote down the amended equation immediately afterwards. Thus, it seems safe to assume that Ornstein was familiar with the note, and it was this knowledge that motivated him to credit Einstein with the germinal idea for converting Eq.~(1) into a stochastic differential equation. 

\subsection{Langevin's 1908 paper}

Langevin accepted the validity of the equipartition principle by writing 
\begin{equation}
m\overline{v^2}={RT\over N},
\end{equation}
where a bar denotes `an average extended over a large number of identical particles', and the other symbols have their usual meanings. Using $\mu$ and $a$ instead of Einstein's $k$ and $P$, respectively, he wrote the equation of motion in the form 
\begin{equation}
m{d^2 x\over d t^2}= - 6\pi\mu a {dx\over dt} + X,
\end{equation}
and added immediately afterwards: ``About the complementary force $X$, we know that it is indifferently positive and negative and that its magnitude is such that it maintains the agitation of the particle, which the viscous resistance would stop without it''. It is to be noted that he says `we know' while Einstein wrote `we must assume'. 

Langevin multiplied Eq.~(3) by $x$, and arrived, after some elementary manipulations at the following equation:
\begin{equation}
{m\over 2}{d^2 x^2\over d t^2}-mv^2 = -3\pi\mu a {dx^2\over dt}+Xx .
\end{equation}
Next, he took an ensemble average, argued that `the average value of the term $Xx$ is evidently null by reason of the irregularity of the complementary forces $X$', replaced the average $m\overline{v^2}$ by its equipartition value, introduced the symbol
\begin{equation}
z=\overline{dx^2\over dt}={d\,\overline{x^2}\over dt},
\end{equation}
and arrived finally at his key equation:
\begin{equation}
{m\over 2}{d z\over d t}+ 3\pi\mu az={RT\over N}.
\end{equation}
He was seeking after an expression for $z$, whose stationary limit was to be equated to $2D$; this limit, to be denoted by $z_{\infty}$, emerges as soon as one sets $dz_{\infty}/dt=0$. The result agrees with Langevin, who took a slightly longer route and found
\begin{equation}
D={RT\over 6\pi\mu a N},
\end{equation}
thereby duplicating Einstein's result.

\subsection{An assessment of Langevin's paper}
It is worth remarking first that Langevin's use of symbols and his further manipulations are reminiscent of the steps choreographed by Clausius in a paper where he derived the virial theorem.${}^{14}$ Clausius used the symbol $X$ for the deterministic force when he wrote Newton's equation of motion for a particle (moving in a non-resistive medium), $m(d^2x/dt^2)=X$, and made use of the following identity:
\begin{equation}
2x{d^2 x\over d t^2}+2\left ({dx\over dt}\right )^2 = {d^2x^2\over dt^2}.
\end{equation}
Neither the conception of Eq.~(3) nor its transformation into a differential equation for $x^2$ can be called truly original. Be that as it may, it was Langevin who actually wrote down Eq.~(3) and---regardless of whose brainchild it was---he eventually became its eponym. 

My main concern is with Langevin's justification for setting $\overline{xX}=0$. Many later authors have simply taken his word for the vanishing of this average. The reception accorded to his analysis may be summed up by quoting from a recent review:${}^{15}$ ``Langevin's derivation, however, was spectacularly simple and direct compared to the others (this is probably why it is Langevin's derivation rather than Einstein's that is usually found in modern textbooks)''. Among those who have endorsed Langevin's logic, one may mention Feynman,${}^{16}$ Reif,${}^{17}$, Heer${}^{18}$  and the authors${}^{19}$ of a recent treatise devoted to the Langevin equation. 

I quote Feynman (who uses the symbol $F$ for $X$): 
\begin{quotation}
Now what about $x$ times the force? If the particle happens to have gone a certain distance $x$, then, since the irregular force is {\it completely\/} irregular and does not know where the particle started from, the next impulse can be in any direction relative to $x$. If $x$ is positive, there is no reason why the average force should also be in that direction. It is just as likely to be one way as the other. The bombardment forces are not driving it in a definite direction. So the average value of $x$ times $F$ is zero. [Italics in the original]
\end{quotation}

This line of reasoning would lead one to infer, as has indeed been done, that $\overline{v X}$ and $\overline{xv}$ also vanish. Let us deal with $\overline{v X}$ first. J. D. van der Waals Jr.${}^{20}$ multiplied Eq.~(3) by $v$, took an ensemble average, and set $\overline{v X}=0$ to find
\begin{equation}
{m\over 2}{d\,\overline{v^2}\over dt}=-6\pi\eta a\overline{v^2}.
\end{equation}
Van der Waals knew that Eq.~(9) was unacceptable, since it implied that $\overline{v^2}$ must decay to zero, contravening the equipartition principle. There are two ways to resolve the difficulty: one may either scrap the supposition $\overline{v X}=0$, or discard Eq.~(3) itself. Unfortunately, van der Waals made the latter choice, and spent some effort in formulating his own theory of Brownian movement, using a different stochastic equation. Though van der Waals did not succeed in winning support for his point of view, he contributed indirectly by engaging Ornstein in a fierce debate. In the course of rebutting van der Waals, Ornstein and Burger (O\&B) had to concede that Eq.~(3) is an approximation, albeit an excellent one.$^7$ The limitations of Langevin's equation, which are easily perceived (see below), ought to be emphasized even when it is introduced in an elementary course.

To notice the untenability of the assumption $\overline{xv}=0$, made by Heer,${}^{18}$  one only has to note that since
\begin{equation}
{d \,\overline{x^2}\over d t}= 2 \,\overline{xv}\quad  \hbox{(identically)},
\end{equation}
the vanishing of $\overline{xv}$ implies the vanishing of $D$ itself. 

\subsection{The treachery of physical intuition}

It seems devilishly difficult to divine, on the basis of physical intuition alone, the statistical independence of two random variables in a dynamical problem.${}^{21}$ Let no one forget that Maxwell assumed, in his first attempt to deduce the law of distribution of molecular velocities, that the distribution of the velocity component along a particular cartesian axis was independent of the values along the other two axes.${}^{22}$ According to Chapman and Cowling,${}^{22}$  ``it would be natural to suppose that these components are not independent''. Seven years later, Maxwell repudiated this assumption, and made a second attempt,${}^{23}$ which too proved to be unsatisfactory when scrutinized by Boltzmann. After attributing the proof presented in their book to Lorentz (who improved Boltzmann's exposition) Chapman and Cowling comment: ``This proof also is open to some objection, because of the assumption $\ldots$ that there is no correlation between the velocity and the position of a molecule''. 

Let us return now to Langevin, who accepted that $z$ eventually attains a stationary value. After averaging Eq.~(4), he should have seen that only one conclusion was justified: the value of $\overline{xX}$ is (or becomes, at long times) a constant that cannot be smaller than $-RT/N$. Had he allowed for the possibility  $\overline{xX}\neq 0$, he would have found, instead of Eq.~(6), 
\begin{equation}
{m\over 2}{d z\over d t}+ 3\pi\mu az={RT\over N}+\overline{xX},
\end{equation}
from which one can only deduce that
\begin{equation}
D={z_\infty \over 2}={1\over 6\pi\mu a}\left ( {RT\over N} + \overline{xX}\right ).
\end{equation}
Since Langevin's aim was to validate Eq.~(7), the assumption $\overline{xX}=0$ reduces his demonstration to a tautology. 

The foregoing arguments amount to little more than an elaboration of a footnote in which O\&B, who used the symbol $\Delta$ for the mean-squared displacement, stated: ``From the formula (1) [Eq.~(3) here] we can deduce the relation $\overline{\Delta^2}=bt$ if we introduce suppositions, it is however impossible to find the value of $b$, without penetrating into the mechanism of the Brownian motion''.${}^{7}$

\subsection{Statistical properties of the random acceleration }

I find it hard to evade the conclusion that Langevin was tempted into making an error that has dodged detection on account of its seductive simplicity. Still, I will now adopt the exegesis, offered by O\&B as well as Manoliu and Kittel,${}^{24}$ that Langevin's pithy phrase `indifferently positive and negative' is to be interpreted as follows: it implies not only that $\overline{X}=0$ but also a supplementary relation, concerning the {\it autocorrelation\/} of $X$, from which one can deduce the vanishing of the {\it cross-correlation\/} $\overline{xX}$.  But these authors also noted the necessity of proving the vanishing of $\overline {xX}$ from the second relation. It is imperative to do so because, as Manoliu and Kittel point out,${}^{24}$ physical intuition might lead one to conclude that $\overline {xX}\neq 0$; after all, an encyclopaedia entry,${}^{25}$  cited by them, does warn the reader: ``The mean random force vanishes, but the mean product of the random force into the displacement [in three dimensions] is equal to $-3kT$; the frequent assertion that this product vanishes is mistaken''.  Since proving the relation $\overline {xX}\neq 0$ entails, even if one takes the short cut devised here, a good deal of preparatory work, the route taken by Langevin, which has been praised for its brevity and simplicity, turns into a wearisome trek (from the student's point of view), making Einstein's treatment look like a jaunt---unless the teacher is prepared to replace the assertion `it is self-evident that $\overline {xX}=0$' by the assurance `it can be shown that $\overline {xX}=0$'.

Enough has now been  said to convince the reader that Langevin's derivation of the mean squared displacement must be rectified, or replaced by a cogent alternative. All the ingredients for accomplishing the latter task were assembled by de Haas-Lorentz, but she stopped short, as did Langevin before her, of going beyond the stationary limit. It was Ornstein who, actually worked out, using his own approach, the complete expression in 1917. The prospects of adding something novel to this topic are infinitesimal indeed, but a repackaging of the material scattered in Ornstein's many publications seems worthwhile.  I wish to present a slightly different and shorter account than appears in the 1930 article${}^{26}$  of Uhlenbeck and Ornstein (U\&O). As a preliminary, however, I wish to recall some statements made by Ornstein in his first paper.

\section{Ornstein's 1917 paper}

Ornstein began his first paper on Brownian motion by citing the works of Smoluchowski and Burger; referring to the thesis of de Haas-Lorentz, later published as a monograph,${}^{10}$  he remarked:
\begin{quotation}
It is worth observing that the way different averages depend on time may be calculated from the results obtained by Mrs. de Haas-Lorentz by a slightly more careful transition of the limit than was necessary for the object she had put herself (viz. the determination of the stationary condition). First, I want to determine these averages by a new method \dots 
\end{quotation}

The relation used by de Haas-Lorentz and mentioned by Ornstein has the following form (though they use different symbols):
\begin{equation}
m{d v\over d t}= - \zeta v + X,
\end{equation}
which has some advantages over the form chosen by Langevin: The use, on the right-hand side, of the symbol $\zeta$ rather than $6\pi\eta a$ saves some writing, besides extending the applicability of the equation to non-spherical particles; the introduction of the particle velocity $v$ also simplifies the resulting formulas, but the main reason for the change is to emphasize that velocity, rather than the position, is the quantity of greater interest here. Upon dividing Eq.~(13) by $m$, and setting 
\begin{equation}
\alpha \equiv {\zeta\over m}\quad \hbox{and}\quad  A\equiv {X\over m}.
\end{equation}
one gets the still neater form
\begin{equation}
{d v\over d t}= - \alpha v + A .
\end{equation}

Ornstein carried out a formal integration of Eq.~(15) to get 
\begin{equation}
v=v_0 e^{-\alpha t}+ e^{-\alpha t}\int_{0}^{t}\!e^{\alpha t_1}A(t_1) dt_1 ,\quad v_0\equiv v(0).
\end{equation}
The fact that $A(t_1)$ on the right-hand side is a random function did not worry him unduly, since he knew that he would attribute physical significance only to averaged quantities.  It is important to recall at this stage that, in the works published during 1917--1930, Ornstein calculated an average over those molecules which start, at time $t=0$, with a definite velocity (say $v_0$); I will refer to an average of this type as a $v_0$-average, and distinguish it from an overall average (the quantity considered by Langevin) by appending the symbol $v_0$ to the bar signifying a particular average. A $v_0$-average can itself be averaged over all initial velocities (distributed according to the Maxwell-Boltzmann law). If one's aim is to obtain only the overall average, it often proves more convenient to bypass the $v_0$-averaging. For the sake of completeness, let us take note of a mean value that will be needed shortly. If we take a $v_0$-average of Eq.~(15), we arrive, after setting $\overline{A\,}^{v_0}=0$, at the equation
\begin{equation}
{d \,\overline{v\,}^{v_0}\over d t}= - \alpha \,\overline{v\,}^{v_0},
\end{equation}
whose solution is
\begin{equation}
\overline{v\,}^{v_0}= v_0 e^{-\alpha t}.
\end{equation}
This is, of course, the same result as would be obtained by taking a $v_0$-average of Eq.~(16). 
 
At this point, I would like to part company---at least temporarily---with Ornstein and turn to the problem of calculating the quantity $\overline{xA}$. 

\section{Emending the Langevin analysis}
Since we are dealing with a system that is in thermal equilibrium, we can immediately state
\begin{equation}
\overline {v}= 0,\quad\quad
\overline {v^2}={RT\over Nm}=\beta, \hbox{ say.}
\end{equation}
Our task is to deduce the statistical properties of $A$, {\it using only Eqs.~(15) and (19)\/}. If we take an ensemble average of Eq.~(16), we get, after setting $\overline {v_0}= 0$ on the right-hand side, 
\begin{equation}
\overline {v}=  \int_{0}^{t}\!e^{\alpha t_1}\overline {A(t_1)} dt_1 =0,
\end{equation}
from which follows the relation $\overline{A}=0$. 

Averaging the equation found by multiplying Eq.~(15) by $2v$ gives
\begin{equation}
{d \,\overline{v^2}\over d t}= - 2\alpha \overline{v^2} + 2\overline{vA}=0,
\end{equation}
which shows, after it is combined with Eq.~(19), that
\begin{equation}
\overline{vA}=\alpha \beta .
\end{equation}
On the other hand, multiplication of Eq.~(16) by $A(t)$ and averaging gives
\begin{equation}
\overline {Av}=  e^{-\alpha t}\int_{0}^{t}\!e^{\alpha t_1}\overline {A(t)A(t_1)} dt_1.
\end{equation}
On equating the right-hand sides of the last two equations one finds
\begin{equation}
\alpha\beta =  e^{-\alpha t}\int_{0}^{t}\!e^{\alpha t_1}\overline {A(t)A(t_1)} dt_1,
\end{equation}
a relation that can only hold if $\overline {A(t)A(t_1)}$ possesses the two properties expressed below; in the first place
$$
\overline{A(t)A(t_1)}=
\cases{0,&if $t_1\neq t$\cr
	\cr
	\infty&if $t_1=t$,\cr}\eqno (\hbox{25a})
$$
and secondly
$$
\int_{0}^{t}\!\overline {A(t)A(t_1)} dt_1=\alpha\beta.\eqno (\hbox{25b})
$$
\setcounter{equation}{25}
We have now completed the preliminaries for finding $\overline{Ax}$. Integration of Eq.~(15) from 0 to $t$ provides a relation connecting $x$, $v$, and $A$:
\begin{equation}
v(t)-v_0=-\alpha \bigl [x(t)-x(0)\bigr ]+\int_{0}^{t}\!A(t_1)dt_1 
\end{equation}
We now multiply Eq.~(26) by $A(t)$ and take an average to find, after some obvious rearrangement,
\begin{equation}
\overline{Ax}={1\over \alpha}\left [\int_{0}^{t}\!A(t)A(t_1)dt_1-\overline{vA}\right ]=0.
\end{equation}
The second equality follows from the fact that each term within the square brackets equals $\alpha \beta$.  After this demonstration, we can justifiably retrace the steps that took Langevin from Eq.~(4) to Eq.~(6), and integrate our version of Eq.~(6) to get
\begin{equation}
{d\,\overline{xv}\over dt}=- \alpha \overline{\,xv\,}+\beta ,
\end{equation}
which can itself integrated to obtain
\begin{equation}
\overline{xv}={\beta\over \alpha}\left (1-e^{-\alpha t}\right ).
\end{equation}
Since $d\overline{\,x^2\,}/dt=2\overline{\,xv\,}$, we finally get, with $x_0\equiv x(0)$,
\begin{equation}
\overline{x^2\,}=x_0^2+{2\beta\over \alpha}\left [ t-{1\over \alpha}\left (1- e^{-\alpha t}\right )\right ],
\end{equation}
a relation that Langevin could have derived but chose not to do so. He was reluctant, as was de Haas-Lorentz a few years later, to go beyond the expression derived by Einstein.  Needless to say, this result was first obtained by Ornstein${}^{5}$ by averaging his $v_0$-average $\overline{\,x^2\,}^{v_0}$ over a Maxwellian distribution of initial velocities.

\subsection{A closer look at correlations}

We see from Eq.~(29) that the correlation of $x$ and $v$, assumed to be zero initially, rises exponentially (with a time constant $1/\alpha$) to its stationary value. On the other hand, the average $\overline{vA}$ turned out to be a constant; this is simply because we have assumed the corresponding time constant to be zero. It is tempting to tinker with the autocorrelation of $A$ and assign it a small but non-zero decay time by setting
\begin{equation}
\overline{A(t_1)A(t_2)}= C \lambda {e^{-\lambda |t_2-t_1|}},\quad C=\hbox{ constant}.
\end{equation}
Should we change our mind, we can always recover the results previously obtained by letting $\lambda \to\infty$ and assigning $C$ an appropriate value. Substituting the right-hand side of the last equation into Eq.~(23) and integrating we get
\begin{equation}
\overline{\,vA\,}={C\lambda\over \lambda+\alpha}\left [1-e^{-(\lambda+\alpha t)}\right ].
\end{equation}
Though this result looks appealing, it should be rejected, when dealing with a system that is in thermal equilibrium, because a time-dependent value of $\overline{\,vA\,}$ violates the equipartition principle: according to Eq.~(21), $\overline{\,vA\,}$ must be a constant, and this constancy is incompatible with any choice that makes the autocorrelation $\overline{\,A(t_1)A(t_2)\,}$ decay at a finite rate. Ornstein and van Wijk (O\&vW), who dealt with overall averages rather than $v_0$-averages, emphasized that the statistical properties of $A$ are determined as soon as one writes down Eqs.~(15) and assumes that the system is in thermal equilibrium.${}^{5}$ Manoliu and Kittel overlooked this restriction when they replaced the white spectrum of $A$ by a Lorentzian;${}^{24}$ it will be pointed out presently that the spectrum of $A$ can be made colored (nonwhite) only by modifying the equation of motion itself. The introduction of Eq.~(31) did not lead to an immediate generalization of our treatment, but it can still serve as a crutch for those students who are unable to jump from Eq.~(24) to Eq.~(25), or feel uncomfortable with an alternative (given below) involving the Dirac delta function; in the next section, it will serve as a part of a scheme devised for generalizing Eq.~(15).

\subsection{A critique of the Langevin equation itself}

Let us set $m=1$ so that we may speak of the force rather than acceleration, and let $w(t)=-\alpha v+ A$ denote the total force acting on the particle. O\&B agreed with van der Waals that for $t=0$, $\overline{w(t)\,}^{v_0}=0$. Besides it is evident, they remarked, that for $t$ infinite the average value of $w(t)$ undergoes no influence from $v_0$ and therefore must be zero. They noted that the course of $\overline{w(t)\,}^{v_0}=0$ must be such that it starts from zero, attains a maximum (or minimum) and then falls to zero again. On the other hand, Eqs.~(15) and (18) show that 
\begin{equation}
\overline{w\,}^{v_0}=-\alpha \overline{v\,}^{v_0}+ \overline{A\,}^{v_0}=-\alpha e^{-\alpha t}v_0.
\end{equation}
\vspace*{-5truemm}
\begin{figure}[htb]
\begin{center}
\includegraphics[height=6cm,width=9cm,angle=0]{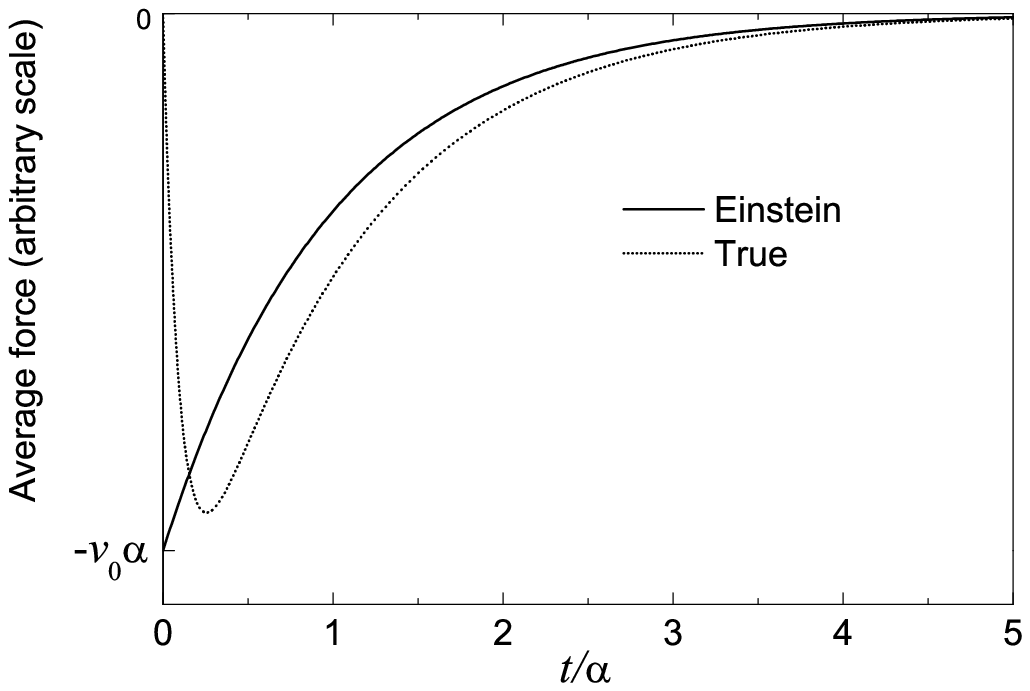}
\end{center}
\end{figure}
\vspace*{-5 truemm}

\noindent{\small Fig.~1. Plots showing the time dependence of the average force acting on a Brownian particle that started with a velocity $v_0>0$. The solid curve depicts the idealized behavior predicted by Eq.~(15), whereas the dotted curve is obtained by plotting the right-hand side of Eq.~(34) with $\lambda = 10\alpha$ and $C=12$.}

\vspace*{5 truemm}

Assuming that $v_0>0$, they drew two curves, one of which, labelled by them as `Einstein', is a plot of the right-hand side of Eq.~(33), and the other (which they call the true curve) is a schematic plot of the behaviour described above. I have taken the liberty of converting their qualitative, freehand figure to a quantitative plot, by using the following expression
\begin{equation}
\overline{w\,}^{v_0}= C{\bigl ( e^{-\lambda t}- e^{-\alpha t}\bigr )\over \lambda - \alpha },\quad (C=\hbox{constant}).
\end{equation}
The reasons for choosing the above form will be explained shortly; at present, it is enough to note that the resulting curve resembles that drawn by O\&B, and to proceed by quoting their comments on the shapes of the curves:
\begin{quotation}
For $t=0$ the line, which represents this course deviates from the true curve. The important agreement existing between Einstein's theory and the experiment now makes us presume, that the true $w(t)$-$t$ curve and the curve according to Einstein only deviate from each other for short times after the departure of the particle with the velocity $v_0$ so that the maximum in the true curve lies close to $t=0$, and that from this maximum onward it descends pretty well exponentially according to Einstein's curve. It goes without saying that these are only assumptions, which a calculation of the true $\overline{w(t)\,}^{v_0}$ curve must prove from the molecular theory. We are however of the opinion that it is worthwhile to point to this possible interpretation of Einstein's master-stroke in the theory of the Brownian motion.
\end{quotation}

It is remarkable that O\&B recognized the limitations of Eq.~(15) as far back as 1918, and even pointed out the nature of the refinment that was needed. A generalization of Langevin's equation in the form
\begin{equation}
{d v\over d t}= - \int_{-\infty}^{t}\alpha (t-t^\prime) v(t^\prime ) + A(t) .
\end{equation}
came almost fifty years later.${}^{27}$ It can be shown that $\alpha (t-t^\prime)$ is proportional to the autocorrelation of $A$, which explains the choice made in Eq.~(34): if we take a $v_0$-average of Eq.~(35), and use a zero-order approximation $\overline{v\,}^{v_0}=v_0\exp(-\alpha t)$ on the right-hand side, and apply Eq.~(31), the integral would equal (apart from a multiplicative constant) the right-hand side of Eq.~(34).     

\section{A random walk in velocity space}

Both Langevin and Ornstein began with two claims: each stated that though the desired result(s) could be found by a re-working of an existing analysis, each would rather present a derivation of his own.  Ornstein mentioned that the time dependent averages can be calculated from the results obtained by de Haas-Lorentz,${}^{10}$  who credited the gist of the method used by her to a 1910 article (not concerned with Brownian motion) by Einstein and Hopf.${}^{28}$  She integrated Eq.~(15) from $t=(N-1)\tau$ to $t=(N-1)\tau$, thereby converting it into a difference equation:
$$
v_N-v_{N-1}= -\alpha  v_{N-1}\tau +  I_{N-1} ,\eqno \hbox{(36a)}
$$
where
$$
I_{N-1}= \int_{(N-1)\tau}^{N\tau}A(t_1) dt_1 \eqno \hbox{(36b)}
$$
\setcounter{equation}{36}
The time interval $\tau$ is so short that $v$ does not change substantially during this interval, but the fluctuations in $A$ are supposed to be extremely rapid even on this short time scale. With this background, we can already state two statistical properties of the random term on the right-hand side of Eq.~(36a):
\begin{equation}
\overline{I_k\,}^{v_0}=0
\end{equation}
\begin{equation}
\overline{I_jI_k}^{v_0}=\delta_{ij}\Pi_1, \quad
\delta_{ij} = \cases{1,&if $j=k$\cr
	\cr
	0,&otherwise.\cr}
\end{equation}

We have arrived at Eq.~(36a) by discretizing Eq.~(15), but an alternative point of view, adopted by Gunther and Weaver,${}^{29}$  is worth a mention: One can regard Eq.~(36a) as the basic equation in its own right, $I_0$, $I_1$, $\cdots$ being a sequence of mutually independent random variables; a great virtue of Eq.~(36a) is its amenability to Monte Carlo simulation.${}^{29,30}$ Regardless of one's interpretation, Eq.~(36a) can be immediately understood as describing {\it a random walk in velocity space\/}, and its solution requires only summation of geometric series, as demonstrated first by de Haas-Lorentz; her contribution has been underestimated, partly because she herself called it the method of Einstein and Hopf, but mainly because she did not extract the full time-dependence from her expressions. 

It follows from iteration of Eq.~(36a) that $v_N$ can be arranged as a sum. Since this section is supposed to provide a discrete variant of Ornstein's approach, we will express the sum so as to highlight the correspondence between Eq.~(16) and its discrete counterpart shown below:
\begin{equation}
v_N= v_0\rho^N + \rho^{N} \sum_{k=1}^{N}I_{k-1}\rho^{-k}  ,
\end{equation}
where
\begin{equation}
\rho\equiv (1-\alpha \tau).
\end{equation}
Let us now take a $v_0$-average of Eq.~(39), and arrive, after using Eq.~(37),
at the relation
\begin{equation}
\overline{v_N\,}^{v_0}=v_0\rho^N
\end{equation}
We would expect, on letting $N\to \infty$ and $\tau \to 0$, to recover the results found by Ornstein if we make the identifications $N\tau =t$, so that $v_N=v(t)$, $x_N=x(t)$, and so on. Since we are going to let $\tau\to 0$, it is permissible to write
\begin{equation}
\rho = e^{-\alpha \tau},\quad\hbox{or}\quad\rho^N=e^{-\alpha t}. 
\end{equation}
We have now established that, as $\tau\to 0$, one gets
\begin{equation}
\overline{v\,}^{v_0}= v_0e^{-\alpha t},
\end{equation}
a result that coincides with that stated in Eq.~(18).

Upon squaring-and-averaging, Eq.~(39), and making use of Eqs.~(37) and (38), we are left with
\begin{eqnarray}
\overline{v_N^2}^{v_0}&=&v_0^2\rho^{2N}+\rho^{2N}\Pi_1 \sum_{k=0}^{N-1}\rho^{-2k} \nonumber\\
&=&v_0^2\rho^{2N}+ \Pi_1 \left [    {\rho^{2N}-1\over 1-\rho^{-2}}   \right ].
\end{eqnarray}
As we let $\tau\to 0$, the first term on the right-hand side goes into $v_0^2e^{-2\alpha t}$, whereas the factor within the square brackets will approach
$$
{1\over 2\alpha\tau}(1-e^{-2\alpha t}).
$$
Since $\overline{v^2}^{v_0}$ must tend to $\beta$ as $t\to\infty$, we choose the limiting value
\begin{equation}
\lim_{\tau\to 0}{\Pi_1 \over \tau } =2\alpha\beta,
\end{equation}
and arrive thereby at the result
\begin{equation}
\overline{v^2}^{v_0}= v_0^2e^{-2\alpha t}+\beta(1-e^{-2\alpha t}),
\end{equation}
derived first by Ornstein.

There is no need to go further since Gunther and Weaver${}^{29}$ have shown how to obtain $\overline{v^2\,}^{v_0}$ as well as $\overline{x^2\,}^{v_0}$, but two comments are still in order. First, they were apparently unaware that the bulk of the calculation had already been done by de Haas-Lorentz; secondly, they arrive at a formula which may be written (when allowance is made for the difference in notation and a typographical error in their paper) as follows:
\begin{eqnarray}
\overline{(x-x_0)^2}^{v_0} &=& {v_0^2\over \alpha^2}(1-e^{-\alpha t})^2 \qquad\qquad\qquad\qquad\qquad\nonumber \\
&& +{2\beta\over \alpha}\Bigl [t -{3-4e^{-\alpha t}+e^{-2\alpha t}\over 2\alpha} \Bigr ].
\end{eqnarray}
At this stage, they only point out that if one sets $v_0^2=\beta$, the expression for the mean-squared displacement given in Eq.~(47) reduces to that `obtained using the classical Langevin equation', namely that given in Eq.~(30). Of course, both expressions pertain to the Langevin equation, and their difference can be traced to different averaging procedures. 

Continuing in the same vein, one can recover all the mean values calculated by U\&O; a verification of this statement will not be offered in these pages.

\section{The delta function}

Ornstein's handling of integrals involving the autocorrelation $\overline{A(t_2)A(t_1)\,}^{v_0}$ reveals an uncanny prescience of an impulse function (some ten years before Dirac${}^{31}$ introduced the delta function). Readers who do not have easy access to Ornstein's 1917 article may examine the two 1927 articles, where the original calculations are repeated. In all these papers, Ornstein encounters the autocorrelation when he calculates $\overline {v^2\,}^{v_0}$; by squaring both sides of Eq.~(16) and performing a $v_0$-average he finds
\begin{eqnarray}
\overline {v^2\,}^{v_0} & = & v_0^2 e^{-2\alpha t}+ \nonumber\\
&&\quad\int_{0}^{t}\!dt_1 \, e^{\alpha t_1}\!\int_{0}^{t}\,e^{\alpha t_2}\overline {A(t_1)A(t_2)\,}^{v_0}  dt_2 .
\end{eqnarray}
At this point he argues that $\overline {A(t_1)A(t_2)\,}^{v_0}$ must be different from zero, since $\overline {v^2\,}^{v_0}$ for $t=\infty$ is positive, but it is clear that the correlation can be appreciable only when the difference between $t_1$ and $t_2$ is small. He introduces $\tau= t_2-t_1$ as a new variable and proceeds as follows:
\begin{eqnarray}
&&\int_{0}^{t}\!dt_1 \, e^{ \alpha t_1}\!\int_{0}^{t}\,e^{\alpha t_2}\overline {A(t_1)A(t_2)\,}^{v_0} dt_2 = \qquad\qquad\qquad\quad\nonumber\\
&&\quad\quad\quad\quad\int_{0}^{t}\!dt_1\,  e^{2\alpha t_1}\!\int \overline {A(t_1)A(t_1+\tau)\,}^{v_0} d\tau , 
\end{eqnarray}
adding that the replacement of $t_1+t_2$ in the exponent by $2t_1$ is justified because only the region around $\tau=0$ contributes to the second integral. Finally, he states that it is easily seen that the second integral is independent of $t_1$ and $t_2$, and is a characteristic constant for the problem at hand, and may be denoted by $a_1$. To a modern reader, it should be evident that Ornstein identified the autocorrelation $\overline {A(t_1)A(t_2)\,}^{v_0}$ with $a_1\delta (t_1-t_2)$. In the 1917 article, he also evaluated $\overline {v(t)A(t)\,}^{v_0}$ through the same reasoning and found that its value to be $a_1/2$. It will be instructive to repeat this calculation using delta calculus. We have
\begin{eqnarray}
\overline{vA\,}^{v_0}&=&e^{ \alpha t}\int_{0}^{t}\, e^{ \alpha t_1}\overline {v(t)A(t_1)\,}^{v_0} dt_1 \nonumber\\
&=&e^{ \alpha t}\int_{0}^{t} \, e^{ \alpha t_1}a_1\delta (t-t_1) dt_1 = {a_1\over 2}
\end{eqnarray}
since
\begin{equation}
\int_{0}^{\epsilon}\delta (x) dx={1\over 2}\int_{-\epsilon}^{\epsilon}\delta (x) dx = {1\over 2}
\end{equation}
Ornstein never used the name {\it delta function\/} but he did use, in two joint publications with Uhlenbeck, the symbol $\delta$. In their 1930 article,${}^{26}$ U\&O mention that 
$$
\left({1\over 4\pi Dt}\right )^{1/2}e^{-(x-x_0)^2/4Dt}
$$
is that solution of the diffusion equation which ``for $t=0$, becomes $\delta (x-x_0)$, when $\delta (x)$  means the function defined by the properties 
\begin{eqnarray*}
\delta (x) &=&0\quad \hbox{for}\quad x\neq 0\\
\int_{-\infty}^{\infty}\delta (x) dx &=& 1 \hbox{"} .
\end{eqnarray*}
Seven years later, Ornstein and Uhlenbeck${}^{32}$ used the symbol $\delta (E-\varepsilon)$ for expressing the condition ``that at $t=0$ all particles have the energy $E$,'' and added ``$\delta (E-\varepsilon)$ is the well-known singular peak function''.   However neither U\&O nor O\&vW  chose to take advantage of the delta calculus, and preferred to carry out their manipulations using the pre-Dirac manoeuvres pioneered by Ornstein; in these two papers, the authors change both variables by setting $u=(t_1+t_2)$ and $w=(t_1-t_2)$. I will not speculate about their preference, but will now raise, and later answer, a different question: Why has Ornstein not been credited with the invention of the delta calculus?

The answer is not far to seek. The second section of the 1927 paper of Dirac is devoted to notation, where he states:
\begin{quotation}
One cannot go far in the development of the theory of $\ldots$ without needing a notation for that function of $\ldots$ $x$ that is equal to zero except when $x$ is small, and whose integral through a range that contains the point $x=0$ is equal to unity. We shall use the symbol $\delta (x)$ to denote this function, {\it i.e.\/}, $\delta (x)$ is defined by
$$
\delta (x)=0\quad \hbox{when}\quad x\neq 0,
$$
and
$$
\int_{-\infty}^{\infty}\!d x \;\delta (x)=1.
$$ 
Strictly, of course,  $\delta (x)$ is not a proper function of $x$, but can be regarded only as a limit of a certain sequence of functions. All the same one can use $\delta (x)$ as though it were a proper function for practically all the purposes of $\ldots$ without getting incorrect results. One can also use the differential coefficients of $\delta (x)$, namely $\delta^\prime (x)$, $\delta^{\prime\prime} (x)\ldots$, which are even more discontinuous and less ``proper'' than $\delta (x)$ itself.

A few elementary properties of these functions will now be given so as not to interrupt the argument later. We can obviously take $\delta (x)=\delta (-x)$, $\delta^\prime (x)=-\delta^\prime (-x)$, etc.  $\ldots$. If $f(x)$ is any regular function $\ldots$, we have
$$
\int_{-\infty}^{\infty}\!f(x) \,\delta (a-x) d x =f(a),
$$ 
so that the operation of multiplying by $\delta (a-x)$ and integrating with respect to $x$ is equivalent to the operation of substituting $a$ for $x$.
\end{quotation}

What we now call Dirac's delta function might well have been named after Ornstein, if the latter had interrupted his argument to introduce, with appropriate fanfare, a new function, spelling out its principal properties once and for all, and giving it a smart name.

\section{The naming of the process and the process of naming}

With the exception of two papers, where Ornstein was the only author, and a third, where he had more than one coauthor, each of his other papers in the period 1917--33 was a joint effort with a single coauthor: Burger, Uhlenbeck, van Wijk and Zernike. Though Uhlenbeck is the first author of the 1930 article, the random process investigated in the article is now called {\it Ornstein-Uhlenbeck process\/}. The person responsible for this reversal of order is, as far as I can see, Doob,${}^{33}$ who must have been under the impression that Ornstein was the first author (as can be verified by examining his list of references); since he refers, throughout his paper, to `Ornstein and Uhlenbeck' it stands to reason that he introduced the name `O.U. process'; subsequent authors have chosen to replace the initials by complete names without reversing their order. Since singling out Uhlenbeck among all the coauthors named above seems (to me) unfair, and including them all is clearly impractical, the choice `Ornstein process' would have been more equitable, but the name `Ornstein-Uhlenbeck process is here to stay.

Ornstein died in 1941.${}^{34}$ Four years later, Ming Chen Wang and Uhlenbeck published their celebrated review article.${}^{35}$ By now, they were unequivocal in calling Eq.~(15) the Langevin equation; they also used the appellation `the Dirac delta function' and used it for specifying the temporal behavior of the autocorrelation of $A$ by stating the relation 
$$
\overline{A(t_1)A(t_2)}=2\alpha\beta \delta(t_1-t_2).
$$

\section{Acknowledgment}

I am very grateful to O. L. J. Gijzeman for his friendly help in acquiring copies of some works by Ornstein and van der Waals, and a copy of ref. 4. 
\vspace {15mm}
\small
\begin{enumerate}
\item{} See, for example, E. Nelson, ``{\it Dynamical Theories of Brownian Motion\/}, (Princeton University Press, 2001), 2nd. edition, p.~45. Posted at the web at: {\tt http://www.math.princeton.edu/$\sim$nelson/books/\\bmotion.pdf} 
\item{}
P. Langevin, ``Sur la th\'{e}orie du mouvement brownien,'' C. R. Acad. Sci. (Paris) {\bf 146}, 530-533 (1908)
\item{}
D. S. Lemons and A. Gythiel, ``Paul Langevin's 1908 paper ``On the Theory of Brownian Motion'' [``Sur la theorie du mouvement brownien, C. R. Acad. Sci. (Paris) 146, 530--533 (1908)],'' Am. J. Phys. {\bf 65}, 1079-1081 (1997). I have followed the translation presented in this article---apart from replacing ``M.'' (the French abbreviation for {\it Monsieur\/}) by ``Mr.'' (the corresponding abbreviation in English for {\it Mister\/}).
\item{}
L. S. Ornstein, {\it A Survey of his Work from 1908 to 1933}, (Utrecht, 1933). This volume contains a complete list of his works (and that of his close associates) published up to 1932 and a few in 1933; each entry contains not only the title of the article but also a brief comment on the contents. For articles published in the Proceedings of the Royal Academy of Amsterdam, references are given to the original as well as the English translation, and the year for the translated work is stated to be the same as the original, and I have followed this practice (which differs from that followed by Ornstein himself). Articles on Brownian motion are listed in Section a (a26, a27, a30, a32, a33, a39, a40, a43); among the contributions to Brownian motion which appeared after the publication of this book, only ref. 5 is of interest in the present context.
\item{}
L. S. Ornstein and W. R. van Wijk, ``On the derivation of distribution functions in problems of Brownian motion,'' Physica {\bf 1}, 235--254 (1933).
\item{}
L. S. Ornstein, ``On the Brownian motion,'' Proc. Royal Acad. Amsterdam  {\bf 21}, 96--108 (1917).
\item{}
L. S. Ornstein and H. C. Burger, ``On the theory of the Brownian motion'', Proc. Royal Acad. Amst. {\bf 21}, 922--931 (1919).
\item{}
L. S. Ornstein, ``Zur Theorie der Brownschen Bewegung f\"ur Systeme, worin mehrere Temperaturen vorkommen,'' Z.  Phys. {\bf 41}, 848--856 (1927).
\item{}
L. S. Ornstein, H. C. Burger, J. Taylor and W. Clarkson, ``The Brownian movement of a galvanometer coil and the influence of temperature of the outer circuit,'' Proc. Royal Soc. A. {\bf 115}, 391--406 (1927).
\item{}
G. L. de Haas-Lorentz, {\it Die Brownsche Bewegung und einige verwandte Erscheinungen\/} (Viewig, Braunschweig, 1913).
\item{}
A. Einstein, {\it Investigations on the theory of the Brownian movement\/} (Dover, New York, 1956). 
\item{}
M. von Smoluchowski, ``Zur kinetischen Theorie der Brownschen Molekularbewegung der Suspsnsionen,'' Ann. Phys. {\bf 21}, 756--780 (1906).
\item{}
A. Einstein, ``Theoretische bemerkungen \"uber die Brownsche Bewegung,'' Z. Elektrochem. {\bf 13}, 41--42 (1907).
\item{}
R. Calusius, ``On a mechanical theorem applicable to heat,'' Phil. Mag. {\bf 40}, 122--127 (1870).
\item{}
M. D. Haw, ``Colloidal suspensions, Brownian motion, molecular reality: a short history,'' J. Phys. Condens. Matter {\bf 14}, 7769--7779 (2002).
\item{}
R. P. Feynman, {\it Lectures on Physics\/} (Addison-Wesley, Reading, Massachusetts, 1963), Vol. 1, Chap.~41.
\item{}
F. Reif, {\it Fundamentals of Statistical and Thermal Physics\/}  (McGraw-Hill, New York, 1965) p.~565.
\item{}
C. V. Heer, {\it Lectures on Physics\/} (Academic, New York, 1972) p.~417.
\item{}
W. T. Coffey, Yu. P. Kalmykov and J. T. Waldron, {\it The Langevin Equation\/}  (World Scientific, Singapore, 2004) p.~14.
\item{}
J. D. van der Waals Jr., ``On the Theory of the Brownian movement, ''Proc. Royal Acad. Amsterdam  {\bf 20}, 1254--1271 (1918).  The author of this article was the son of J. D. van der Waals, well-known for his equation of state and the winner of the 1910 Nobel Prize in Physics. 
\item{}
S. Chapman and T. G. Cowling, {\it The Mathematical Theory of Non-uniform Gases\/} (C.U.P., Cambridge, 1970) p.~73.
\item{}
J. C. Maxwell, ``Illustrations of the dynamical theory of gases,''  Phil. Mag. {\bf 19}, 19--32 (1860).
\item{}
J. C. Maxwell, ``On the dynamical theory of gases,''  Phil. Trans. Royal Soc. {\bf 157}, 49--88 (1867).
\item{}
A. Manoliu and C. Kittel, ``Correlation in the Langevin theory of Brownian motion,'' Am. J. Phys. {\bf 47}, 678 (1979).
\item{}
J. Thewlis (Ed.), {\it Encyclopaedic Dictionary of Physics\/}  (Pergamon, Oxford, 1961) Vol. 1, pp.~511--512.
\item{}
G. E. Uhlenbeck and L. S. Ornstein, ``On the theory of the Brownian motion,'' Phys. Rev. {\bf 36}, 823--841 (1930).
\item{}
R. Kubo, ``Brownian motion and nonequilibrium statistical mechanics,'' Science {\bf 233}, 330--334 (1986).
\item{}
A. Einstein and L. Hopf, ``Statistische Untersuchung der Bewegung eines Resonators in einem Strahlungsfeld,'' Ann. Phys. {\bf 33}, 1105--1115 (1910).
\item{}
L. Gunther and D. L.Weaver, ``Monte Carlo simulation of Brownian motion with viscous drag,'' Am. J. Phys.  {\bf 46}, 543--545 (1978).
\item{}
J. D. Doll and D. L. Freeman, ``Randomly exact methods,''Science {\bf 234}, 1356--1360 (1986).
\item{}
P. A. M. Dirac, ``The physical interpretation of the quantum dynamics,'' Proc. Royal Soc.  A,  {\bf 113}, 621--641 (1927).
\item{}
L. S. Ornstein and G. E. Uhlenbeck, ``Some kinetic problems regarding the motion of neutrons through paraffine,'' Physica  {\bf 4}, 478--486 (1937).
\item{}
J. L. Doob, ``The Brownian movement and stochastic equations,'' Annals of Mathematics,  {\bf 43}, 351--369 (1942).
\item{}
R. C. Mason, ``Leonard Salomon Ornstein,'' Science {\bf 102}, 638--639 (1945).
\item{}
Ming Chen Wang and G. E. Uhlenbeck, ``On the theory of the Brownian motion II,'' Rev. Mod. Phys. {\bf 17}, 323--342 (1945).
\end{enumerate}

\end{document}